\documentclass[12pt]{article}
\usepackage[cp1251]{inputenc}
\usepackage{graphicx}
\usepackage{indentfirst}
\usepackage{afterpage}
\usepackage{longtable}

\def\gee{ \, \lower 1mm\hbox{$\,{\buildrel > \over{\scriptstyle\scriptstyle\sim} }\displaystyle \,$}}
\def\lee{ \, \lower 1mm\hbox{$\,{\buildrel < \over{\scriptstyle\scriptstyle\sim} }\displaystyle \,$}}
\def\|{\partial}

\def\varkappa {{\scriptstyle\partial}\! e}

\let\c=\centerline

\let\b=\baselineskip
\begin{document}
\headheight 1.50true cm \headsep  0.7true cm \righthyphenmin=2


\large


\begin{center}
 \textbf{The relationship between the neutral hydrogen and dark mass in the galaxies}
\end{center}

\c{ Zasov A.V.$^1$, Terekhova N.A.$^1$}

\c{\it $^1$M.V. Lomonosov Moscow State University, 
Sternberg Astronomical Institute}

\bigskip
\textbf{Abstract} - Starting from  Bosma' (1981) paper, it was demonstrated by different authors that the observed shape of rotation curves of many spiral galaxies can be explained if to assume that the radial density distribution of the dark matter is correlated with the distribution of $HI$: the column densities of the dark matter and $HI$ are proportional. We show that this correlation is obviously to be an artifact and can be explained by assuming that the gas surface density is approximately equal or in general proportional to the critical density for the local gravitational stability of gaseous layer.

\c{\bf 1. Introduction}
A rotation curve of every single galaxy can be interpreted in terms of the different assumptions about the relative mass and the spatial distribution of visible and dark matter ($DM$) in the galaxy. Moreover, if the rotation of the inner regions of discs can usually be attributed to the mass of visible matter only (Palunas, Williams 2002), then the presence of the dark matter usually seems unavoidable beyond 2-3 radial disc scale length $R_0$. The $DM$ mass needed to explain kinematic properties of galaxies appears to be comparable with the observed galaxy components inside of the optical radius, although $DM$ distribution geometry is known a priori. Theoretically, $DM$ can be located both in the disc and in the quasi-spherical halo, however the gravitational stability requirements and the situation with elliptical galaxies which possess $DM$ in the absence of a disc make the dark halo case more reasonable. The existence of the massive dark halo agrees both with the  mass-to-light ($M/L$) ratios for stellar discs found from photometry (de Blok et al. 2008; Zasov et al. 2011) and the numerical cosmological simulations of galaxies formation in the field of dark halos. However, the question of the presence of essential mass of $DM$ inside galactic discs still remains the subject of active discussions.  In the framework of the existing models of formation of galaxies, a certain percentage of non-baryonic $DM$ may be concentrated in the disc along with the baryonic matter, for example, as the result of tidal disruption by the disc field of satellites containing $DM$ (see, e.g., Read et al. 2009). Note, however, that the $DM$ density inside Galaxy disc (in the solar vicinity) appears to be much lower than the visible components (see, e.g., Garbari et al. 2012).

It is surprising that in many cases the shape of the extended rotation curve may be reproduced without any dark halo if to assume that all (or at least the main part) of $DM$ is concentrated inside the disc, not in the halo, and the dark matter surface density changes with radius proportional to the $HI$ density, although it can exceed it several times (more than tenfold in some cases). For the first time it was mentioned by Bosma (1981). Later this conclusion was confirmed by other authors for different galaxy types (Karignan 1985; Karignan, Pushe 1990; Sancisi 1999; Jobin, Karignan 1990; Hoekstra et al. 2001; Hessman, Ziebart 2011; Swaters et al. 2012). However, there appears a serious problem with gravitation stability of the ``heavy'' gaseous disc, if the main part of $DM$ is concentrated near to its plane (Elmegreen 1995; Revaz et al. 2009; Terekhova 2012). Hence, there still remains a necessity to introduce a dark halo, unless it is assumed that the $DM$ disc is too thick.

The possible interpretation of rotation curves of different galaxy types using the ``heavy'' gas model (baryonic scaling model) was considered by Swaters et al. (2012) for 43 spiral galaxies of different types. It was confirmed, that the observed rotation curves can satisfactorily be reproduced well within the ``heavy'' gas model for most of galaxies, by introducing some scale factor $\eta$, defined as the ratio of the disc density connected with $HI$ to the observed density of neutral hydrogen: $\eta = \Sigma_{gas+DM}/ \Sigma_{HI}$. The scaling factor $\eta$ is proved to be different for the different galaxies characterized by large spread of values. Remarkably, that for the late-type spiral galaxies with higher amount of $HI$ the scale factor $\eta$ appears to be lower and the correlation between the observable and model curves is better, than for early-type galaxies.

The link between $DM$ and neutral hydrogen densities, if it really exists, is difficult to justify physically. Pfenniger et al. (1994) assumed the presence of directly
non-observable medium, which possibly consists of small clouds of molecular hydrogen, too cold to be detected by emission. But even in this case there is  no obvious reason why do non-observable gas and $HI$ have similar radial distributions. So the question remains: whether the masses and spatial distributions of $HI$ and $DM$ are really physically connected?

\medskip
\c{\bf 2. A radial profile of the gas density and the stability of gas layer}
 \medskip
In this paper we argue that the proportionality of the densities of dark matter and $HI$ is an artifact. It is worth noting that the surface brightness (density) of stellar disc falls much steeper (exponentially) than the surface density of $HI$.  Hence, if to assume the existence of additional disc component with the gently sloping radial profile in the galaxy model, one may explain the high rotation velocity at large $R$ without any halo.

This idea was first mentioned in general form by Terekhova (2012). Here it is discussed more detail.

Let's consider the  azimuthally averaged density profile of gaseous layer. Excluding the central disc areas which aren't so essential in our case, the interstellar gas is usually dominated by atomic hydrogen $HI$. So we can assume that its surface density is proportional to the total gas density, i.e. $\Sigma_{HI}\sim \Sigma_{gas}$. The gravitational stability criterion limits the maximum surface density of gas, so it cannot be very high for the observed dispersion velocities of turbulent gas $c_{gas}\sim$ 8-12 km/s. Otherwise, the development of instability would lead to the raise of non-circular motions and, as a consequence, to the change of star formation rate and energy release, until the interstellar gaseous layer does not evolved to the quasi-equilibrium gravitationally stable state.

The observed distribution of the gas surface density $\Sigma_{gas}(R)$ in galaxies can be compared with the distribution of critical density $\Sigma_{crit}\approx (1/Q_T)\cdot(\pi
c_{gas}\Sigma_{gas}/\varkappa)$ corresponding to marginal stability for gravitational perturbations in a disc plane (see, e.g., Martin, Kennicutt 2001; Wong, Blitz 2002; Boussier et al. 2003; Leroy et al. 2008 and references to earlier papers).
 Here $Q_T$ is the Toomre's stability parameter which is equal to unit for purely radial perturbations of a thin disc, and $\varkappa$ is the epicyclical frequency equal to $\sqrt{2}V/R$ for the ``flat'' rotation curve. A comparison of $\Sigma_{crit}(R)$ with $\Sigma_{gas}(R)$ shows that in a wide interval of $R$ the density $\Sigma_{gas}$ usually either remains close to $\Sigma_{crit}$ (within the factor 1.5-2), or it changes approximately parallel with $\Sigma_{crit}$, although there are some exceptions. On the periphery of the discs this ratio decreases as a rule. However, it should be taken into account that gas velocity dispersion is usually accepted to be constant when evaluating $\Sigma_{crit}$, while it actually slowly decreases at  large distances $R$  (de Blok et al. 2008), making the outer disc regions to be closer to the threshold of the gravitational stability. Thus, the $\Sigma_{gas}/\Sigma_{crit}$ ratio really appears to change slowly in the wide interval of $R$ in many cases.

For non-spiral discy galaxies the situation may be different. In dwarf irregular galaxies the $\Sigma_{gas}/\Sigma_{crit}$ ratio is as a rule less than 1 being almost constant with $R$ or (more often) passing through maximum followed by decrease to the disc periphery (van Zee et al. 1997; Hunter et al. 2011). For the early type galaxies the condition $\Sigma_{gas}/\Sigma_{crit}=const$ can be applied to a lesser extent (Noordermeer et al. 2005);  usually they have $\Sigma_{gas}/\Sigma_{crit}<1$  both in the central and in the outer disc regions. Hence, it is not surprise that the ``heavy'' gas model poorly explains the rotation curves of galaxies of early types.

For more accurate estimations of $\Sigma_{crit}$ the destabilizing influence of the stellar disc should be taken into account. The calculations of the gaseous layer stability considering the stellar disc gravitation (following Rafikov 2001) were carried out for some spiral galaxies by Leroy et al. (2008). They demonstrated that the $\Sigma_{gas}/\Sigma_{crit}$ ratio changes more slightly with $R$ than that obtained without the accounting for stellar disc. The gaseous discs appear to be stable, but, as a rule, localize close up to the stability threshold (see Fig. 9 from Leroy et al. 2008). Thus, radial distribution of $HI$ in galaxies is usually connected with the rotation disc velocity (by the condition of stability) which does not take into account in the ``heavy'' gas model.

As the simplest example, let's consider a galaxy with the ``flat'' rotation curve where the gaseous layer density is close to the critical one. In this case $\varkappa=\sqrt{2}V/R$ and if we neglect the changing of $c_{gas}$ and $Q_T$ with radius, the gas density  $\Sigma_{gas}\sim\Sigma_{crit }\sim 1/R$. The disc with such density distribution is known as Mestel' disc. This disc really corresponds to the ``flat'' rotation curve, as in the case of the pseudo-isothermal halo. So it is to be expected that in the ``heavy'' gas model, where the gaseous density is multiplied by some fitting coefficient, the rotation curve plateau of such galaxy will be well reproduced.

The less obvious situation is in the cases of no plateau on the rotation velocity curve which instead continues to rise or decreases at large $R$. Nevertheless, even in these cases the rotation curve fits satisfactory to the ``heavy'' gas model. It's easy to show that if the galaxy rotation curve increases or decreases with $R$, the same behavior may be expected for the ``heavy'' gas disc model if to assume $\Sigma_{gas}\approx \Sigma_{crit}\sim \varkappa$, neglecting the $Q_T$ variation with $R$. To illustrate it, consider the smoothed rotation curves taken from the THINGS review (de Blok et al. 2008) for the galaxies NGC 2976 (the growing one) and NGC 4736 (the declining one). In Figs. 1 and 2 they are compared by the shape (scale-free) with the rotation curves expected for the disc with $\Sigma_{disc}\sim \varkappa$  density distribution. The Casertano's (1983) method (GR3 program, the author is A.N. Burlak) for the finite disc half-width (taken as 0.5 kpc to be specific) was used for the construction of the model rotation curve which conforms to the radial density profile. As it follows from Figs. 1 and 2, the condition $\Sigma_{gas}\approx \Sigma_{crit}$ explains qualitatively the shape of the rotation curves even if there is no plateau.

In general, to model the rotation curves of real galaxies, a mass of the observed stellar disc should be taken into account, although its contribution decreases along $R$ due to the disc density decreasing.  It is worth considering the applicability of the ``heavy'' gas model for the concrete late-type galaxies (Table 1) taking into account their stellar discs. The presented sample contains both normal surface brightness (HSB) galaxies (their rotation curves for the different galaxy components and the HI radial profiles were taken form the THINGS review (de Blok at al. 2008)) and low surface brightness (LSB) galaxies. The latter as a rule possess the higher $M_{HI}/M_{stellar disc}$ ratios, and, if the invisible matter in galaxies is connected with $HI$, its influence on the disc dynamics should be particularly noticeable. The data for the rotation curves and their stellar and gas components for LSB galaxies parallel with their $HI$ radial profiles were taken from de Blok et al. (2001). The $M/L$ ratio of stellar disc for every galaxy was accepted to be constant, as in the case of the HSB galaxies.

For every galaxy we calculated the rotation curves using the two-component model of galaxy, which includes a stellar disc with constant $M/L$ ratio (taken from the sources cited above), and a ``heavy'' gas layer with the density   $\eta$-times higher than the directly-measured $HI$ density. Parameter $\eta$ was chosen from the best fitting between the model and observed rotation curves.

Results of modeling allow to divide the galaxies into three subgroups by the concordance of their rotation curves with the ``heavy'' gas model (Table 1). Here we conditionally assume that the model rotation curve generally corresponds to the observed (smoothed) one if a discrepancy between them does not exceed 30\% within some range of radial distances given in Table 1. For galaxies of the first group a satisfactory fitting takes place within the radius $\sim 1-2R_0$ only (here $R_0$ is the radial luminosity scale length in B-band). For the second group the fitting of the model and the observed rotation curves is successful within the larger range $R=1-4R_0$. Galaxies of the third group do not agree at all with the ``heavy'' gas model. The ranges of the stability parameters for gas layer for a given interval of $R$ with and without the accounting of stellar disc ($Q_T=Q_{stars+gas}$ and $Q_T=Q_{gas}$ correspondingly) for all three groups are also presented in Table 1. Values of $Q_{stars+gas}$ were taken from Leroy et al. (2008), where they used the indirect estimates of stellar velocity dispersion obtained for equilibrium and constant-width stellar disc. For the other galaxies, the stability parameters were calculated assuming the gas velocity dispersion $c_{gas}=11$ km/s.

As it follows from Table 1, among those 13 galaxies (groups I and II) where the ``heavy'' gas model does not conflict too much with observations, there are only three galaxies (NGC 2403, NGC 6946 and NGC 7793), a stability parameter of which varies significantly (by 30\% or more) in the specified interval of $R$.  Note that $Q_{stars+gas}$ estimates  given in Table 1 are enclosed within the more narrow intervals than $Q_{gas}$.  When the ``heavy'' gas model reveals a poor fitting (Group III), the stability parameters vary within very wide ranges. This suggests that the distribution of the azimuthally averaged gas density in these galaxies is far from being proportional to $\Sigma_{crit}(R)$.  Apparently, it explains the discrepancy between  ``heavy'' gas models and the observed rotation curves.

As an illustration, below we consider the rotation curves for galaxies of each group.  In Fig. 3 for the galaxy NGC 925 (the model with ``heavy'' gas is suitable only for the interior of the disc) are given the observed rotation curve (circles), gas and stellar components of the rotation curve from de Blok et al. (2008) (dotted and dashed lines, correspondingly), and the rotation curve for the ``heavy'' disc model (triangles). A satisfactory agreement between the observed and model curves holds for $R<$10 kpc. Fig. 4 shows the change of $Q_{gas}$ and $Q_{stars+gas}$ with radius for this galaxy (according to Leroy et al. 2008). The $Q_{stars+gas}$  parameter, taking into account the stellar disc,  changes less than at 30\% within  $2R_0\sim 8$ kpc. In Figs. 5 and 6 similar relations are presented for the DDO 154 galaxy of Group II,  for which the ``heavy'' gas model fits to the observations in the range $R=$0.8-3.2 kpc. The LSB galaxy F568-3 is presented in Figs. 7 and 8. The ``heavy'' gas model for this galaxy poorly describes the observed rotation curve, and the $Q_{gas}$ parameter varies  within the very wide limits.

\medskip

\c{\bf 3. Summary and conclusion}
 \medskip
Radial profile of gas surface density is not independent on the galaxy rotation curve shape. In many cases (but not always) gaseous layer density changes along $R$ in compliance with its marginal gravitational stability requirement. This fact is ignored in the discussion of the ``heavy'' gas models, where the unobserved matter (dark mass) is assumed to be inside the disc, and its density to be proportional to the density of $HI$. If the gas surface density $\Sigma_{HI}$ changes as $R^{-1}$ within a rather wide interval of $R$, the rotation curve component provided by a ``heavy'' gas would come to ``plateau'', imitating the pseudo-isothermal dark halo contribution to the rotation curve (after the proper choice of the fitting coefficient $\eta$).  In some galaxies $\Sigma_{HI}$ falls down faster than $R^{-1}$ at large radial distances, hence a higher value of $\eta$ coefficient is needed to reproduce the real rotation curve in the ``heavy'' gas model. This is consistent with the fact, that for galaxies where the rotation curve is traced farther from the center the coefficients $\eta$ are usually higher (Swaters et al. 2012). If the rotation curve in the outer disc does not come to plateau, and continues to increase, or conversely, decreases with $R$, the analogous behavior will reveal a component of the rotation curve belonging to marginally stable gaseous layer. Therefore in these cases the observed rotation curve can also be reproduced by multiplying the observed $HI$ density at some fitting coefficient and ignoring a dark halo, although this procedure has no physical sense. We show that in those galaxies, where ``heavy'' gas model fails to reproduce the shape of the observed rotation curve, the observed gaseous layer density does not follow the marginal stability condition.

A general conclusion is that the connection between the dark matter and $HI$ densities is only apparent, being caused by the fact that the gas density distribution in many galaxies depends on the shape of the  rotation curve, following the condition of marginal gravitational stability of gaseous layer.

Note added in proof:
A similar conclusion was independently obtained in a recent paper by Meurer et al. (2013).

\medskip

\c{\bf Acknowledgments }
 \medskip
This work was supported by grants RFBR 12-02-00685 and 11-02-12247-ofi.

\bigskip
\pagebreak
 
\pagebreak


{\tiny
\begin{longtable}{|c|c|c|c|}
\caption{The ranges of the stability parameters $Q_{gas}, Q_{stars+gas}$ for galaxies considered.}\label{Tabl1} \\
\hline {\it 1} & {\it 2}& {\it 3} &{\it 4}\\
\hline Galaxy & $R_0$, kpc & $Q_{gas}$ &$Q_{stars+gas}$
\\\hline

\hline
\multicolumn{4}{|c|}{Group I} \\
NGC 925 &   $4.1^{(1)}$ &   2.5-3.8 & 1.5-1.7 \\
NGC 2976 & $0.9^{(1)}$ &    8.9-12.2 & 1.6- 2.0 \\
NGC 3109 &  $4.0^{(2)}$ &   2.9-3.3 & - \\
\multicolumn{4}{|c|}{Group II} \\
NGC 1560 & $1.3^{(3)}$ & 1.1-1.6 & - \\
NGC 2366 & $1.59^{(4)}$ &   1.3-1.9 & - \\
NGC 2403 & $1.6^{(1)}$ & 4.4-10.4 & 1.7-2.7 \\
NGC 3741 & $0.16^{(5)}$ & 3.3-3.8   & - \\
NGC 6946 & $2.5^{(1)}$ & 1.8-7.03 & 1.2-1.9 \\
NGC 7793 & $1.3^{(1)}$ & 3.99-8.71 & 1.5-2.1 \\
IC 2574 &   $2.1^{(4)}$ & 1.5-2.7 & 1.1-2.3 \\
DDO 154  & $0.8^{(1)}$ &    6.8-7.2 & - \\
F563-1 &    $3.2^{(7)}$ & 4.4-5.7 & - \\
F571-8 & $2.8^{(7)}$ & 2-2.7 &  - \\
\multicolumn{4}{|c|}{Group III (model does not explain the rotation curve)} \\
NGC 3521 &  $2.7^{(1)}$ &   1.8-7.4 &   1.1-2.3 \\
NGC 5055 &  $3.2^{(1)}$ & 0.3-3 & 0.9-2.9 \\
NGC 5585 &  $1.4^{(6)}$ & 4.1-7.8   & - \\
F568-V1 &   $3.8^{(7)}$ & 4.6-11.3 &    - \\
F568-3 & $3.0^{(7)}$ & 3.7-6.3 &    - \\
F574-1 &    $3.5^{(7)}$ & 5-9.8 &   - \\
F579-V1 & $3.1^{(7)}$ & 6.2-11.3 &  - \\
F583-1 & $1.2^{(7)}$ & 3.7-7.6 &    - \\

\hline
\end{longtable}
Note: values of the radial scalelength of the disc $R_0$ in $B$-band
are taken from: (1) - Leroy et al. (2008), (2) - Jobin, Carignan (1990), (3)
- Broeils (1992), (4) - Oh et al. (2008), (5) - de Blok et al. (1996),
(6) - Cote et al. (1991), (7) - de Blok et al. (2001).

\pagebreak

\begin{figure} [h!]

\includegraphics[width=11cm,keepaspectratio]{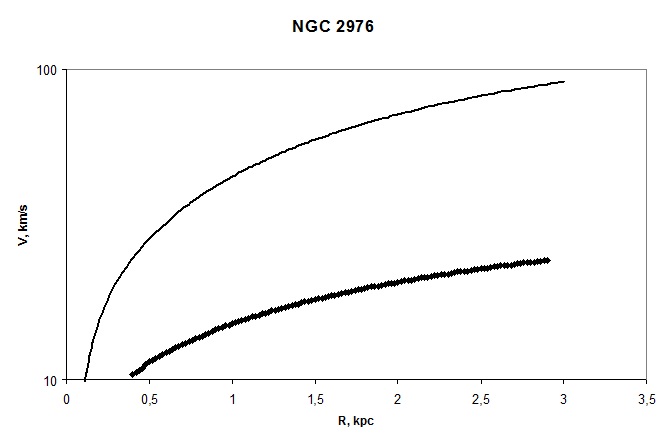}
\caption{The example of growing rotation curve is shown as a thin line (a smoothed curve for NGC 2976 following de Blok et al. 2008);  thick line  is the rotation curve for a single disc with $\Sigma_{disc}\sim \varkappa$ (a shift along the logarithmic vertical axis is arbitrary).}
\end{figure}
\begin{figure} [h!]

\includegraphics[width=11cm,keepaspectratio]{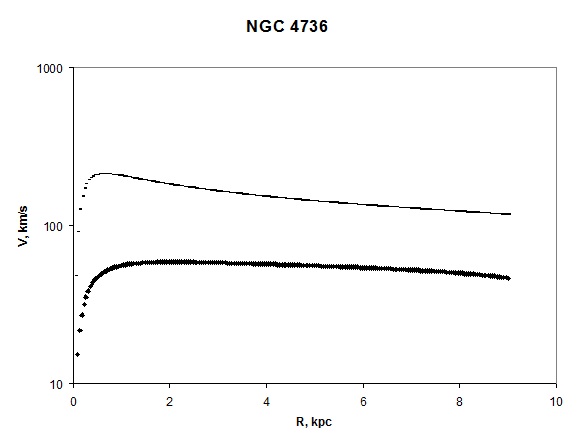}
\caption{Curves similar to those at Fig. 1 for galaxy with decreasing velocity of rotation NGC 4736.}
\end{figure}
\begin{figure} [h!]

\includegraphics[width=11cm,keepaspectratio]{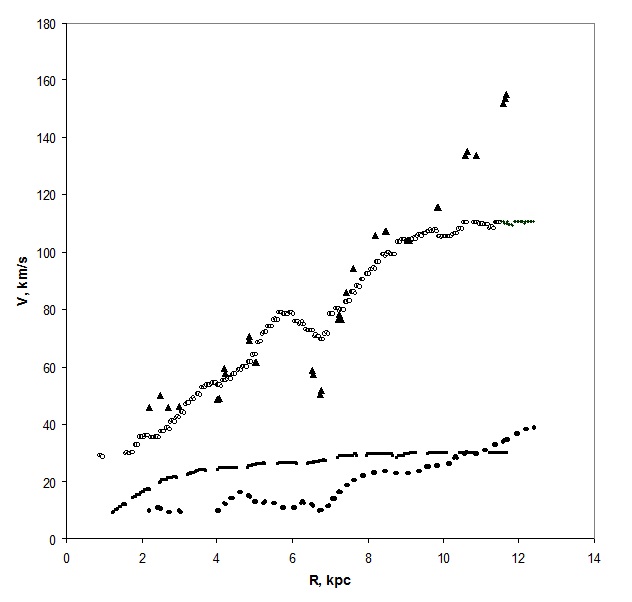}
\caption{Rotation curve for NGC 925 and its components (following de Blok et al. 2008). Dashed and dotted lines are for stellar and gaseous components of rotation curve correspondingly. Circles mark the observed rotation curve. Triangles -- rotation curve in the model with a ``heavy'' gas.}
\end{figure}
\begin{figure} [h!]

\includegraphics[width=11cm,keepaspectratio]{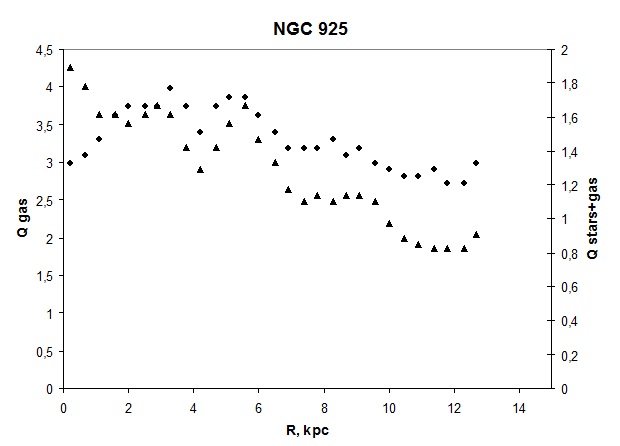}
\caption{NGC 925: parameters of gravitational stability for the gas layer without (triangles) and with (points) the accounting of  stellar disc (see the digitization along the right axis).}
\end{figure}
\begin{figure} [h!]

\includegraphics[width=11cm,keepaspectratio]{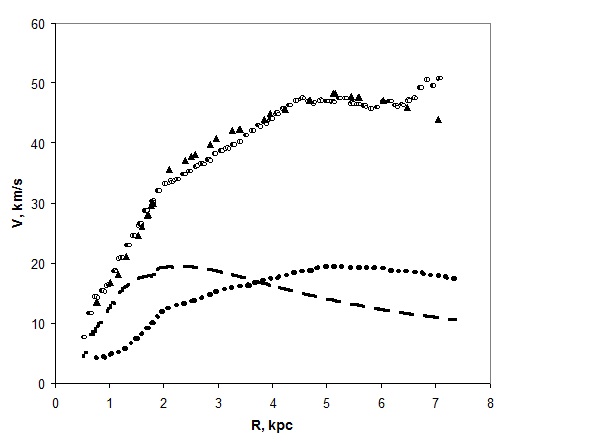}
\caption{Rotation curve for DDO 154. Designations are similar to those at Fig. 3. The shape of the curve is well described by a model with a ``heavy'' gas.}
\end{figure}
\begin{figure} [h!]

\includegraphics[width=11cm,keepaspectratio]{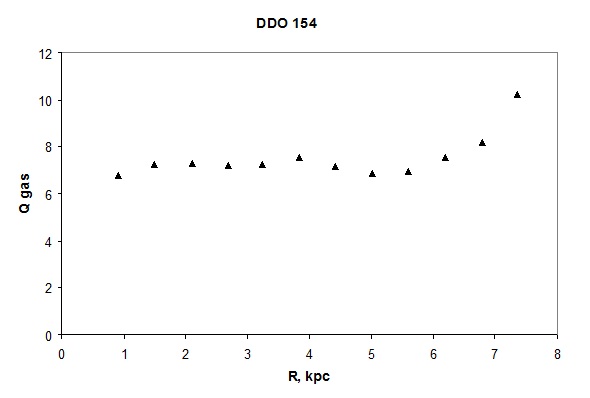}
\caption{DDO 154: a parameter of gravitational stability $Q_{gas}$ for the gas layer.}
\end{figure}
\begin{figure} [h!]

\includegraphics[width=11cm,keepaspectratio]{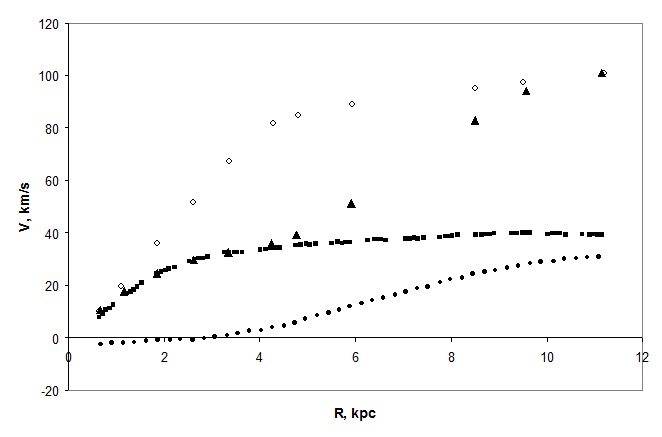}
\caption{Rotation curve and its components for F568-3 (following de Blok et al. 2001). Designations are similar to those at Fig. 3. The shape of the curve is badly described by  model with a ``heavy'' gas (triangles). }
\end{figure}
\begin{figure} [h!]

\includegraphics[width=11cm,keepaspectratio]{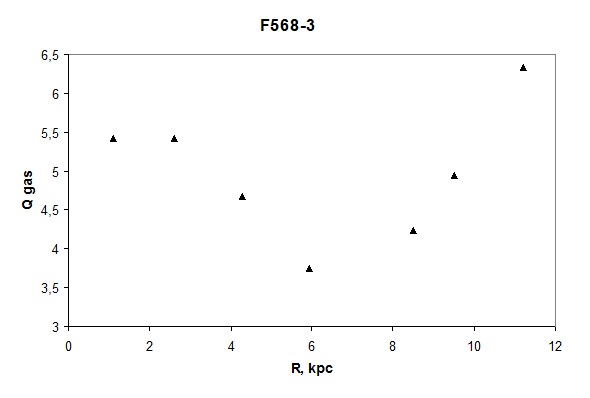}
\caption{F568-3: a parameter of gravitational stability $Q_{gas}$ for the gas layer. It is characterized by a large spread of values.}
\end{figure}

\end{document}